\documentclass[aps,preprintnumbers,amsmath,amssymb]{revtex4}

\usepackage{float}
\usepackage{graphics,epsfig}
\usepackage{graphicx}
\usepackage{dcolumn}
\usepackage{bm}

\begin{document}
\baselineskip=0.8 cm
\title{{\bf A Note on Self-gravitating Radiation in AdS Spacetime}}
\author{Zhong-Hua Li}\affiliation{Department of Physics, China West Normal University, Nanchong
637002, China}

\author{Bin Hu}
\email{hubin@itp.ac.cn} \affiliation{ Institute of Theoretical
Physics, Chinese Academy of Sciences, P.O.Box 2735, Beijing 100190,
China}

\author{Rong-Gen Cai}
\email{cairg@itp.ac.cn} \affiliation{Institute of Theoretical
Physics, Chinese Academy of Sciences, P.O.Box 2735, Beijing 100190,
China}

\vspace*{0.2cm}
\begin{abstract}
\baselineskip=0.6 cm
\begin{center}
{\bf Abstract}
\end{center}
Recently Vaganov [arXiv:0707.0864] and Hammersley [arXiv:0707.0961]
investigated independently the equilibrium self-gravitation
radiation in higher ($d\geq 4)$ dimensional, spherically symmetric
anti-de Sitter space. It was found that in $ 4 \le d \le 10$, there
exist locally stable radiation configurations all the way up to a
maximum red-shifted temperature, above which there are no solutions;
there is also a maximum mass and maximum entropy configuration
occurring at a higher central density than the maximal temperature
configuration.  Beyond their peaks the temperature, mass and entropy
undergo an infinite series of damped oscillations, which indicates
the configurations in this range are unstable.  In $d \geq 11$, the
temperature, mass and entropy of the self-gravitating configuration
are monotonic functions of the central energy density, asymptoting
to their maxima as the central density goes to infinity. In this
note we investigate the equilibrium self-gravitating radiation in
higher dimensional, plane symmetric anti-de Sitter space. We find
that there exist essential differences from the spherically
symmetric case: In each dimension ($d\geq 4$), there are maximal
mass (density), maximal entropy (density) and maximal temperature
configurations; they do not appear at the same central energy
density; the oscillation behavior appearing in the spherically
symmetric case, does not happen in this case; and the mass
(density), as a function of the central energy density, increases
first and reaches its maximum at a certain central energy density
and then decreases monotonically in $ 4\le d \le 7$, while in $d
\geq 8$, besides the maximum, the mass (density) of the equilibrium
configuration has a minimum: the mass (density) first increases and
reaches its maximum, then decreases to its minimum and then
increases to its asymptotic value monotonically. The reason causing
the difference is discussed.

\end{abstract}

\maketitle
\newpage

\section{Introduction}

Over the past decade, a lot of attention has been focused on anti-de
Sitter (AdS) space and relevant physics due to the conjecture of
AdS/CFT correspondence~\cite{Mald}, which says that string theory/M
theory on an AdS space (times a compact manifold) is dual to a
strong coupling conformal field theory (CFT) residing on the
boundary of the AdS space. According to the AdS/CFT correspondence,
Witten~\cite{Witten} argued that thermodynamics of black holes in
AdS space can be identified to that of dual strong coupling CFTs.
Therefore one could study thermodynamics and phase structure of
strong coupling CFTs by investigating thermodynamics and phase
structure of AdS black holes. It is well-known that thermodynamics
of AdS black holes is quite different from that of their
counterparts in asymptotically flat space. For the AdS Schwarzschild
black hole, there is a minimal temperature, below which there is no
black hole solution, above which there are two black hole solutions
with a same temperature. The large AdS Schwarzschild black hole is
thermodynamically stable with positive heat capacity, while the
small AdS Schwarzschild black hole is thermodynamically unstable
with negative heat capacity like Schwarzschild black holes. Below
the minimal temperature, only does the self-gravitating radiation
(thermal AdS gas) solution exist. Between the large AdS black hole
and thermal AdS gas, there is a one-order phase
transition~\cite{HP}, named the Hawking-Page phase transition.
According to the AdS/CFT correspondence, this phase transition can
be identified as the confinement/deconfinemnet phase transition in
the dual gauge theory~\cite{Witten}. Therefore it is of interest to
study self-gravitating radiation configuration in AdS space.

Sorkin, Wald and Zhang~\cite{SWZ} have studied the equilibrium
configurations of self-gravitating radiation in a spherical box of
radius $R$ in asymptotically flat space. It was found that for a
locally stable configuration, the total gravitational mass of
radiation obeys the inequality $M<\mu_{max} R$, where
$\mu_{max}=0.246$. In AdS space, one needs not any unphysical
perfectly reflecting walls at finite radius. The rising
gravitational potential in AdS space, plus natural boundary
conditions at infinity, acts to confine whatever is
inside~\cite{HP}. In \cite{PP} Page and Philips examined the
self-gravitating configuration of radiation in four dimensional AdS
space. The configuration can be labeled as its mass, entropy and
temperature versus the central density. They found that there exist
locally stable radiation configurations all the way up to a maximum
red-shifted temperature, above which there are no solutions; there
is also a maximum mass and maximum entropy configuration occurring
at a higher central density than the maximal temperature
configuration.  Beyond their peaks the temperature, mass and entropy
undergo an infinite series of damped oscillations, which indicates
the configurations in this range are unstable. The self-gravitating
radiation in five dimensional AdS space has been studied in
\cite{HLR} (see also \cite{Hemm}) with similar conclusions.

Recently, Hammersley~\cite{Hamm} and Vaganov~\cite{Vaga}
independently discussed the self-gravitating radiation
configurations in higher dimensional AdS spacetimes. They found that
in the case of $4 \le d \le 10$, the situation is qualitatively
similar to the case in four dimensions, while with $ d \geq 11$, the
oscillation behavior disappears. Namely, there is a critical
dimension, $d=11$ (very close, but not exact), beyond which, the
temperature, mass and entropy of the self-gravitating configuration
are monotonic functions of the central energy density, asymptoting
to their maxima as the central density goes to infinity. For related
discussions in asymptotically flat space see also \cite{Chav}.

On the other hand, let us note that in asymptotically AdS space,
black hole horizon can be not only a positive constant curvature
surface, but also a zero~\cite{zero} or negative constant curvature
surface~\cite{negative}. Such black holes are called topological
black holes in the literature. In particular, it was found that
while the Hawking-Page phase transition exists for AdS black holes
with spherically symmetric horizon in $d\geq 4$, it does not appear
for a Ricci flat (plane symmetric) or hyperbolic horizon~\cite{Topo}
(For Ricci flat black holes, if one of the spatial directions is
compact, the Hawking-Page transition can happen due to the existence
of AdS soliton, see \cite{CKW} and references therein).

Note that Refs.~\cite{Hamm} and \cite{Vaga} considered the
spherically symmetric self-gravitating radiation configuration. It
is therefore of great interest to see whether there is also a
critical dimension in the plane symmetric ($k=0$) self-gravitating
radiation in AdS space, and to see whether there is any essential
difference between the spherical symmetric case and the plane
symmetric case. The organization of the paper is as follows. In the
next section, we give a general formulism to describe the
self-gravitating radiation in AdS space and  study the plane
symmetric case. The numerical results are given in Sec.~III. Section
IV is devoted to the conclusions.

\section{Self-gravitating radiation in AdS space}

Consider a $d$-dimensional asymptotically AdS space with metric
\begin{equation}
\label{2eq1}
 ds^2= -e^{2\delta(r)} h(r) dt^2 + h^{-1}(r)dr^2 + r^2 \gamma_{ij}
 dx^idx^j,
 \end{equation}
 where $\delta$ and $h$ are two functions of the radial coordinate $r$,
 and $\gamma_{ij}$ is the metric of a $(d-2)$-dimensional Einstein manifold with constant
 scalar curvature $(d-2)(d-3)k$. Without loss of generality, one can
 take $k=1$, $0$ and $-1$, respectively. We take the gauge
 $\lim_{r\to \infty} \delta (r) =0$, and rewrite the metric function
 $h(r)$ as
 \begin{equation}
 \label{2eq2}
 h(r) = k+ \frac{r^2}{l^2}- \frac{16\pi G m(r)}{(d-2) \Sigma
 r^{d-3}},
 \end{equation}
 where $l$ denotes the radius of the AdS space with cosmological constant
 $\Lambda=-(d-1)(d-2)/2l^2$, $\Sigma$ is the volume of the Einstein
 manifold,
  and $m(r)$ is the mass function of the
 solution. In our gauge, the total gravitational mass of the solution is
 just
 \begin{equation}
 M = \lim_{r\to \infty} m(r).
 \end{equation}
 The Einstein field equations with the cosmological constant and energy-momentum tensor
 $T_{\mu\nu}$ are
 \begin{equation}
 \label{2eq4}
 R_{\mu\nu}-\frac{1}{2} g_{\mu\nu}R -\frac{(d-1)(d-2)}{2l^2}g_{\mu\nu}= 8\pi G T_{\mu\nu}.
 \end{equation}
 Let us first briefly review the vacuum solution of (\ref{2eq4}). In
 this case, $\delta (r)=0$ and $m(r)$ in (\ref{2eq2}) is just an
 integration constant $M$, which parameterizes the mass of the
 solution. When $k=1$, the solution (\ref{2eq1}) with (\ref{2eq2})
 describes a static, spherically symmetric black hole in AdS
 space. The black hole horizon $r_+$ is determined by $h(r)|_{r=r_+}=0$. Thus the mass
 parameter $M$ can also be expressed in terms of the horizon radius,
 \begin{equation}
 M= \frac{(d-2)\Sigma r_+^{d-3}}{16\pi G} \left(k+
 \frac{r_+^2}{l^2}\right).
 \end{equation}
 Of particular interest is that when $k=0$ or $-1$, the metric is
 also of a black hole cause structure. When $k=0$, namely, the plane symmetric case,
  the black hole
 horizon is $r_+^{d-1}= 16\pi GMl^2/(d-2)\Sigma$. Note that the volume of
 the Ricci flat surface $\gamma_{ij}dx^idx^j$ could be divergent, in this
 case, but
 $M/\Sigma$ is finite and can be viewed as the mass
 density of the solution. Of course, the volume can also be made finite by
 identification.  When $k=-1$, some
 strange features appear. Note that in this case, the volume of the hyperbolic surface
 is also divergent, but one can make it finite by identification and the black hole
 horizon becomes a higher genus closed surface~\cite{negative}.  In the case of $k=-1$,
 one may notice that even when $M=0$,
 the solution also describes a black hole with horizon $r_+=l$,
 although it is locally equivalent to a $d$-dimensional AdS space.
 Sometimes the black hole is called a ``massless" black hole.
 Furthermore, note that when $M>0$ in (\ref{2eq2}), the black hole has only
 a horizon, while $M_{\rm crit} < M<0$, the solution (\ref{2eq2})
 can have two horizons, where
 \begin{equation}
 \label{2eq6}
 M_{\rm crit}=-\frac{(d-2)\Sigma r_{\rm min}^{d-3}}{8\pi G (d-1)}
 \end{equation}
 where $r_{\rm min}^2=(d-3)l^2/(d-1)$, which is the horizon radius for a minimal
 ``negative" mass black hole.  When $M=M_{\rm crit}$, the two horizons
 coincide with each other, and beyond which, the solution
 describes a naked singularity.

 For the black hole solution with any $k$, the Hawking temperature
 is
 \begin{equation}
 \label{2eq7}
 T=\frac{(d-3)}{4\pi r_+}\left(
 k+\frac{d-1}{d-3}\frac{r_+^2}{l^2}\right),
 \end{equation}
 and the entropy associated with black hole horizon still obeys the
 well-known area formula, namely $S=A/4G$, here $A=\Sigma
 r_+^{d-2}$ is the horizon area of the black hole. The free energy
 of the black hole is easy to calculate, which is given
 by~\cite{Topo}
 \begin{equation}
 \label{2eq8}
 F=\frac{\Sigma r_+^{d-3}}{16\pi G}\left(
 k-\frac{r_+^2}{l^2}\right)-M_{\rm crit} \delta_{k,-1}.
 \end{equation}
 Note that here in the case of $k=-1$, one has taken the extremal black hole with
 mass (\ref{2eq6}) as the reference background, while in the cases of $k=1$ and $k=0$,
 the solutions (\ref{2eq2}) with $m=0$ have been considered
 as reference backgrounds. From (\ref{2eq8}), we can see clearly that when $k=0$ and
 $-1$, the free energy is always negative, which means that the
 black hole phase is always dominant, and no Hawking-Page phase
 transition can happen in this case.

 However, when $k=1$, the free energy is negative as $r_+>l$ (large
 black hole), and positive as $r_+<l$ (small black hole), which
 implies that a one-order phase transition occurs here when
 $r_+=l$. This is just the Hawking-Page phase
 transition~\cite{Witten,HP}. The phase transition temperature
 is $ T_{HP}=(d-2)/2\pi l$. This phase transition means that when
 the temperature of thermal gas in AdS space reaches the critical
 temperature $T_{HP}$, the thermal gas will collapse to form a
 black hole in AdS space (in fact the phase transition temperature
 will be higher than $T_{HP}$ since in our calculation we neglect
 the contribution to the free energy of thermal gas in
 AdS~\cite{HP}). Further, we can see from (\ref{2eq7}) that when
 $k=1$, there exists a minimal temperature of black hole, $T_{\rm
 min}= \sqrt{\frac{d-1}{d-3}}/2\pi l$ with corresponding
 horizon radius $r_{\rm min}=l\sqrt{(d-3)/(d-1)} $. Below this temperature,
 there is no black hole solution, only thermal gas configuration
 exists. When $r_+ <r_{\rm min}$, the black hole has a negative heat
 capacity and is local thermodynamically unstable, while as $r_+>r_{\rm min}$,
 the AdS black hole has a positive heat capacity, it is thermodynamically
 stable, and it can be in thermal equilibrium with its surrounding
 Hawking radiation. Note that when the black hole has a horizon in
 the same order as the AdS radius $l$, the black hole mass $M \sim
 l^{d-3}$, that is, the mass is of order 1 in units of AdS radius.

 When $k=0$ and $-1$, one can see that such minimal temperature
 does not exist. The AdS black hole is always thermodynamically stable with
 positive heat capacity for any horizon radius.
  Therefore we see that thermodynamical properties
 of AdS black holes with $k=1$ are quite different from those of
 black holes with $k=0$ and $-1$. One of the aims of this work is to
 see whether there are any essential differences of the
 self-gravitating radiation between the cases $k=1$ and $k=0$.
 As for the case of $k=-1$, one will see shortly that it is
 impossible to have regular self-gravitating radiation
 configuration in that case.

 Now we turn to the self-gravitating radiation configuration in AdS
 space.  In this case, the metric function $\delta(r)$ and the mass
 function $m(r)$ satisfy the following equations~\cite{LC}
 \begin{eqnarray}
 \label{2eq9}
 && \delta'(r)= -\frac{8\pi G r}{(d-2) h(r)}\left(T^t_{\ t}-T^r_{\
 r}\right ), \nonumber \\
 && m'(r)=-\Sigma r^{d-2} T^t_{\ t},
 \end{eqnarray}
 The energy-momentum tensor of the radiation is
 \begin{equation}
 T^{\mu}_{\nu} = {\rm diag} (-\rho, p, p_{\perp},\cdots,
 p_{\perp}),
 \end{equation}
 where $p=p_{\perp}$, and its equation of state obeys $\rho =
 (d-1)p$. Thus the equations in (\ref{2eq9}) reduces to
 \begin{eqnarray}
 \label{2eq11}
&& \delta'(r)= \frac{8\pi G d r}{(d-2)(d-1) h(r)} \rho ,  \\
\label{2eq12}
 && m'(r)=\Sigma r^{d-2} \rho,
 \end{eqnarray}
From the conservation of the energy-momentum tensor,  one can derive
the energy density $\rho(r)$ satisfies the following equation
\begin{equation}
\label{2eq13} \frac{d\rho}{dr}=-\frac{\rho d}{(d-2) r h(r)}\left(
\frac{8\pi G \rho r^2}{d-1} +\frac{(d-2)r^2}{l^2}+ \frac{8\pi G
(d-3) m(r)}{\Sigma r^{d-3}}\right).
\end{equation}
The energy density of radiation can be expressed in terms of the
local temperature in $d$-dimensional curved space as
\begin{equation}
\label{2eq14}
 \rho(r) = a T_{\rm loc}^d(r),
\end{equation}
where $a$ is a dimensional dependent constant describing the degrees
of freedom of the radiation~\cite{Vaga}: $a= (d-1) \pi^{-d/2}
\Gamma(d/2)\zeta(d)g$ and $g= n_B +(1-2^{-(d-1)})n_F$ with $n_B$
being the number of boson spin states and $n_F$ the number of
fermion spin states. The local temperature has a relation to the
proper temperature $T$ as
\begin{equation}
T_{\rm loc}(r)= e^{-\delta(r)} h^{-1/2}(r) T,
\end{equation}
according to the Tolman red-shift relation, here $T$ is a constant.
For the equilibrium configuration, the entropy density of the
radiation is $s= a d T^{d-1}_{\rm loc}/(d-1)$. Thus the total
entropy of the configuration can be obtained by integrating over the
spatial volume
\begin{equation}
\label{2eq16}
 S=\frac{d}{d-1} a^{1/d} \Sigma \int^{\infty}_0
\rho^{(d-1)/d}h^{-1/2} r^{d-2} dr.
\end{equation}
Thus, for a fixed cosmological constant $l^2$, we can get a set of
self-gravitating radiation configurations, labeled by the central
energy density $\rho_c$, by integrating (\ref{2eq12}) and
(\ref{2eq13}), starting from $r=0$ to $r=\infty$ with the boundary
conditions $\rho(0)=\rho_c$ and $m(0)=0$. The equilibrium
temperature can be obtained by (\ref{2eq14}) as~\cite{PP}
\begin{equation}
\label{2eq17}
 T= \lim_{r\to \infty}(a^{-1}\rho(r) h^{d/2})^{1/d},
\end{equation}
where we have used the gauge with $\delta(\infty)=0$. Once obtaining
the proper temperature $T$, to get another metric function
$\delta(r)$, one can use (\ref{2eq14}), rather than
 integrating the equation (\ref{2eq11}),
 \begin{equation}
 \delta(r)= \ln (a^{1/d}\rho^{-1/d}h^{-1/2}T).
 \end{equation}
In this note we are particularly interested in the relation between
the total mass of configuration and the central energy density. To
integrate (\ref{2eq12}) and (\ref{2eq13}), we make a scaling
transformation as follows,
\begin{equation}
r \to l\,r, \ \ \ \rho \to l^{-2} \rho,  \ \ \  m(r) \to l^{d-3}
m(r),
\end{equation}
 so that $r$, $\rho$ and $m$ become dimensionless. In the numerical
 integration, to compare with \cite{Hamm} and \cite{Vaga},
 following~\cite{Hamm}, we will adopt the units $8\pi G=1$ and
 $l=1$, and rescale the mass function as
 $$16\pi G m(r)/(d-2)\Sigma \to m(r).$$
 In that case, the gravitational ``mass" $M$ in the plots in the
 next section in fact is the gravitational mass density, $16\pi G M/(d-2)\Sigma$,
of corresponding self-gravitating configurations. In this way the
gravitational ``mass" $M$  becomes comparable for the two cases of
$k=1$ and $k=0$.

 Further, note that the equations (\ref{2eq12}) and (\ref{2eq13})
 are singular at $r=0$. To avoid this, in the numerical calculations, we will
 start the integration from $r=\epsilon=10^{-5}$ to $r=L=100$ in the case of $k=1$, and
 from $r=\epsilon=10^{-2}$ to $r=L=100$ in
 the case of $k=0$. Obviously, the accuracy of the numerical
 calculations depends on the values of $\epsilon$ and $L$.

\section{Numerical results}

To be more clear to compare the two cases of $k=1$ and $k=0$, let us
first revisit the spherically symmetric case. In this case, $k=1$,
and the volume of $\gamma_{ij}dx^idx^j$ is $\Sigma=
2\pi^{(d-1)/2}/\Gamma[(d-1)/2]$. In $d=4$, Page and
Philips~\cite{PP} found that there is a maximal mass configuration.
For  larger central energy densities, the mass will undergo an
infinite series of damped oscillations, corresponding to unstable
configurations. The case of $d=5$ is similar to the case of
$d=4$~\cite{HLR}.  Hammersley~\cite{Hamm} and Vaganov~\cite{Vaga}
studied the case with higher dimensions. It was found that in $ 4
\le d \le 10$, the behavior is similar to the case of $d=4$ and in
$d \geq 11$, however, the mass, temperature and entropy of the
configuration turn to be monotonic functions of the central energy
density, asymptoting to their maxima as the central density goes to
infinity~\cite{Vaga}. However, the implication of the existence of
the critical dimension $d=11$ is not clear.

\begin{figure}
\includegraphics[width=10cm]{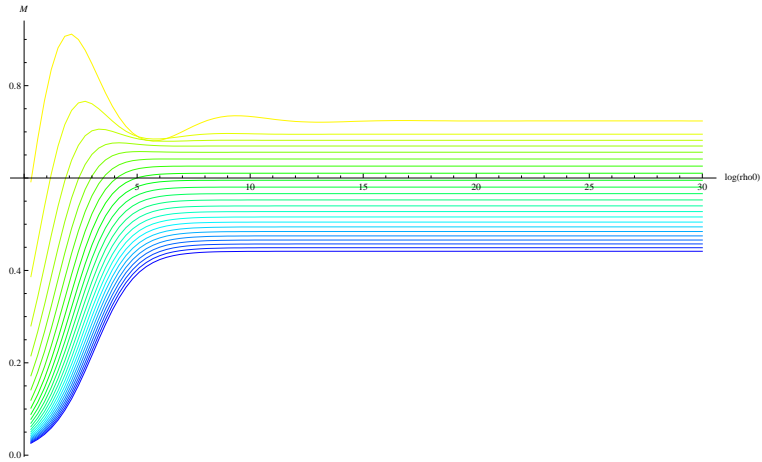}
\caption{\label{massk=1}The spherically symmetric case: The mass of
the self-gravitating radiation configuration versus the central
energy density from spacetime dimension $d=4$ to $d=26$ (from the
yellow to blue curves).}
\end{figure}

\begin{figure}
\includegraphics[width=10cm]{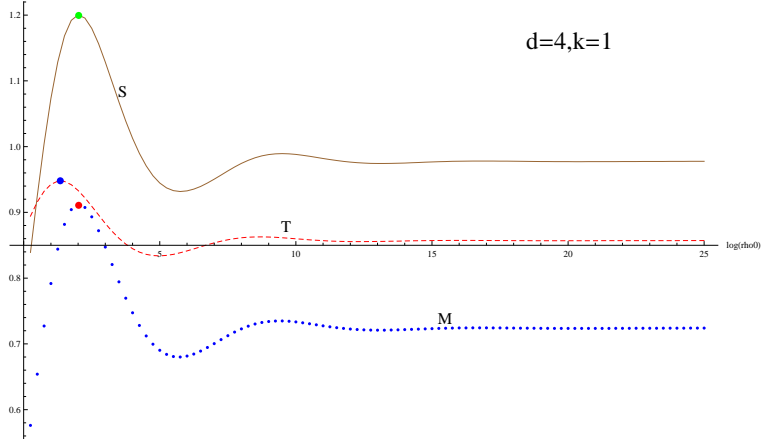}
\caption{\label{mstk=1}The spherically symmetric case: The mass,
entropy and temperature of the self-gravitating radiation
configuration versus the central energy density in the case of
dimension $d=4$.}
\end{figure}

It can be seen from Fig.~\ref{massk=1} that all of configuration
mass is bounded from above by a relatively small value which is of
order one in units of the AdS radius $l$ (note that in the numerical
calculation, $l$ has been set to one). This means in particular that
one of these configurations can at most have a total mass that
equals that of a small AdS Schwarzschild black hole and never that
of a large AdS Schwarzschild black hole. In addition, in $ 4 \le d
\le 10$, the maximal mass configuration and maximal entropy
configuration occur at the same central energy density, while the
maximal temperature configuration appears at a smaller central
energy density~\cite{Hamm,Vaga}. In Fig.~\ref{mstk=1}, we plot the
mass, entropy and temperature of the self-gravitation radiation
configuration in the case of $d=4$, there the oscillation behavior
can be clearly seen.

\begin{figure}
\includegraphics[width=10cm]{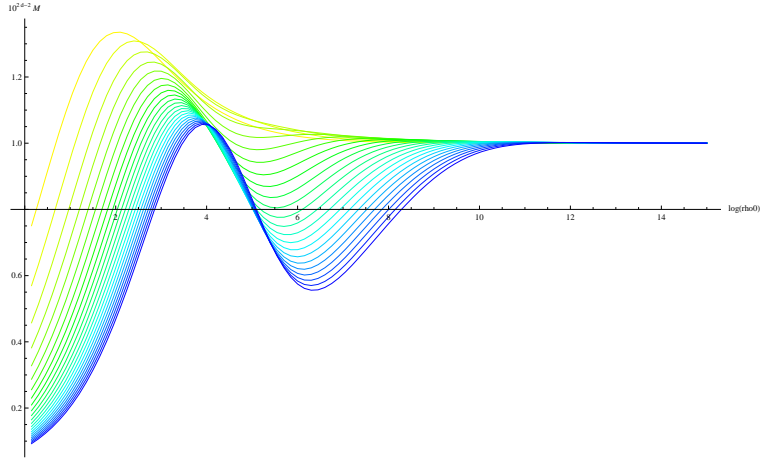}
\caption{\label{massk=0}The plane symmetric case: The mass density
of the self-gravitating radiation configuration versus the central
energy density from spacetime dimension $d=4$ to $d=26$ (from the
yellow to blue curves).}
\end{figure}

\begin{figure}
\includegraphics[width=10cm]{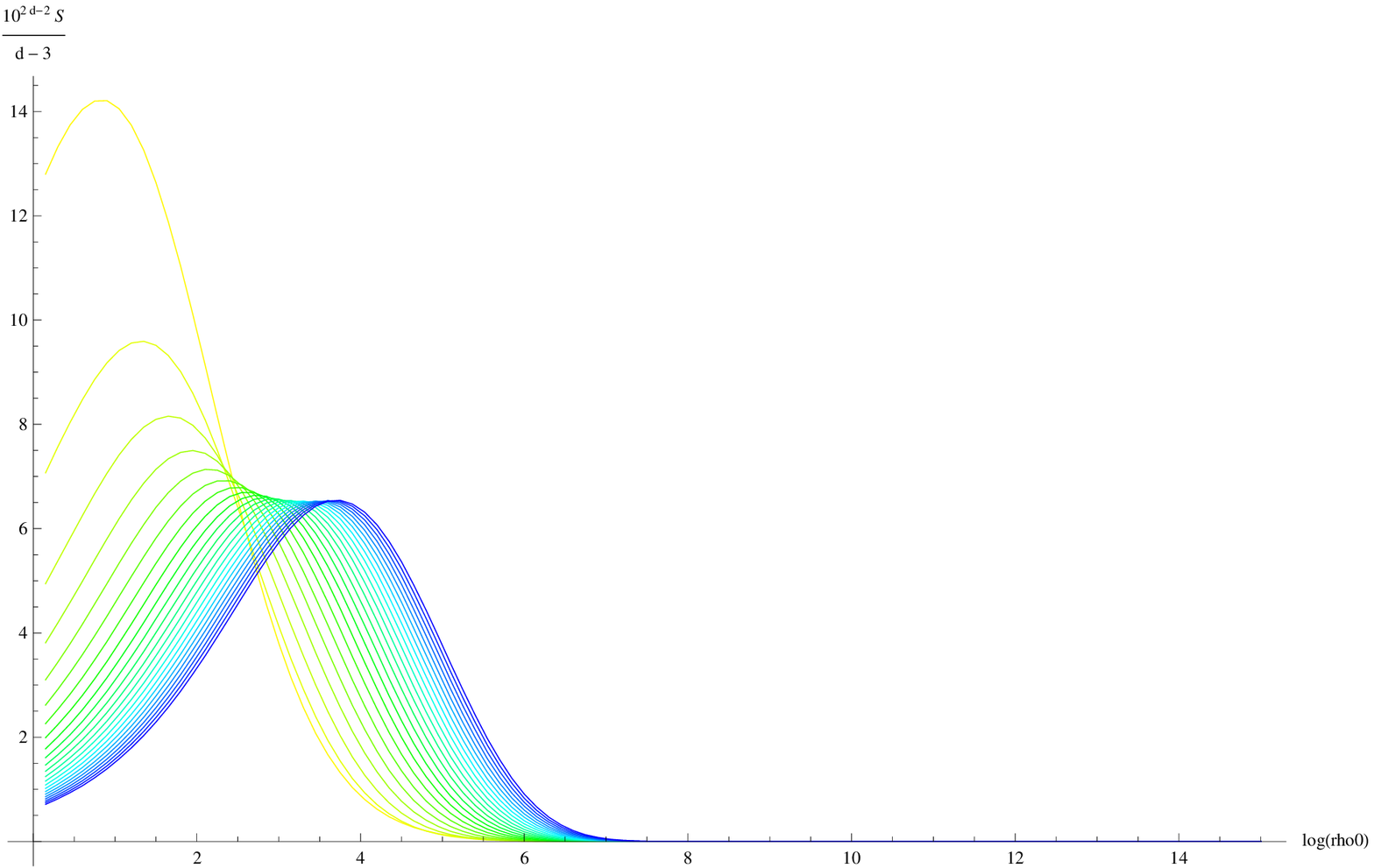}
\caption{\label{entropyk=0}The plane symmetric case: The entropy
density of the self-gravitating radiation configuration versus the
central energy density from spacetime dimension $d=4$ to $d=26$
(from the yellow to blue curves).}
\end{figure}

\begin{figure}
\includegraphics[width=10cm]{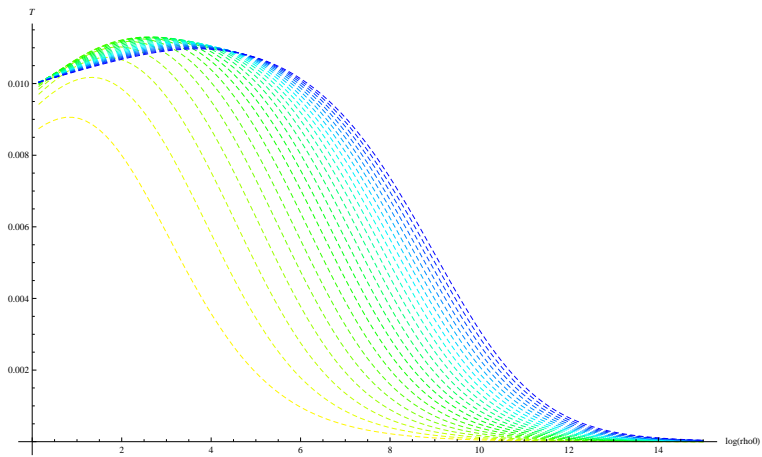}
\caption{\label{temperaturek=0}The plane symmetric case: The
temperature of the self-gravitating radiation configuration versus
the central energy density from spacetime dimension $d=4$ to $d=26$
(from the yellow to blue curves).}
\end{figure}

Now we turn to the plane symmetric case. In that case, we have
$k=0$. In Fig.~\ref{massk=0} we plot the mass density versus the
central energy density. Compared to the spherically symmetric case
$k=1$, some interesting new features appear here. First, the mass
density $M$ is suppressed very much, compared to the case of $k=1$,
although the mass  density is also bounded from above. For example,
the mass density is of order one in the units of AdS radius in the
case of $k=1$, while the mass density
 is in the order $(10^{-6})$ in the case of $k=0$, $d=4$.
Note that in Fig.~\ref{massk=0} we rescale the mass density $M$ with
scale $10^{2d-2}$ in each dimension, which means that as the
spacetime $d$ increases by one, the mass density reduces by two
orders in magnitude.  The suppression of the mass density is due to
the fact that in the case of $k=0$, the black hole is always
thermodynamically stable without any mass limit, while in the case
of $k=1$, the small mass AdS Schwarzschild black hole has a negative
heat capacity and then is thermodynamically unstable. Second, for
each dimension, there is a maximal mass configuration, beyond the
corresponding central energy density, the mass density decreases. We
notice that in $4 \le d \le 7$, the mass density increases and
reaches its maximum at a certain central energy density and then
decreases monotonically to its asymptotic value. In $d \geq 8$, the
mass density increases and reaches its maximum at a certain central
energy density, and then decreases to a minimum at a larger central
energy density, and then increases to its asymptotic value
monotonically. Third, differing from the spherically symmetric case,
beyond the central energy density corresponding to the maximal mass
configuration, there is no any oscillation behavior in the mass
density as a function of the central energy density. In
Figs.~\ref{entropyk=0} and \ref{temperaturek=0} the entropy density
and temperature  of the self-gravitating radiation configuration are
plotted respectively, as functions of the central energy density. In
the plots the entropy is scaled as $10^{2d-2} S
a^{-1/d}/(d-3)\Sigma$ and here $S$ is given by (\ref{2eq16}), the
temperature is scaled as $a^{1/d}T$ and here $T$ is given by
(\ref{2eq17}). Let us mention that although we have rescaled the
mass, entropy and temperature with dimension dependent factors in
Figs.~\ref{massk=0}, \ref{entropyk=0} and \ref{temperaturek=0}, the
shapes of those curves will not change. In addition, note that the
central energy density $\rho_0$ was not rescaled except that we have
taken $l=1$ as the case of $k=1$.

\begin{figure}
\includegraphics[width=10cm]{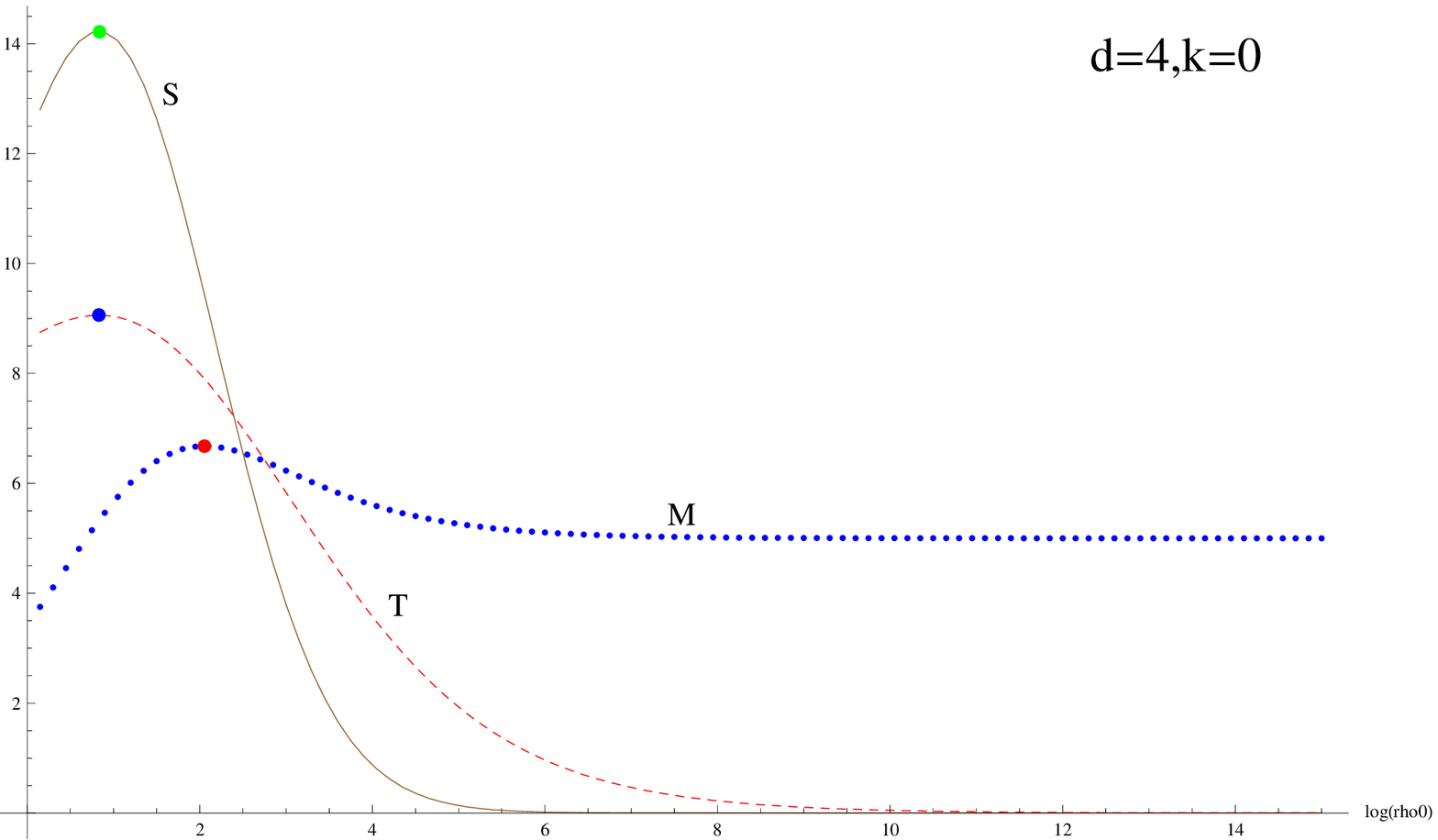}
\caption{\label{mst1d=4}The plane symmetric case: The mass
density, entropy density and the temperature of the
self-gravitating radiation configuration versus the central energy
density in the case of spacetime dimension $d=4$.}
\end{figure}

\begin{figure}
\includegraphics[width=10cm]{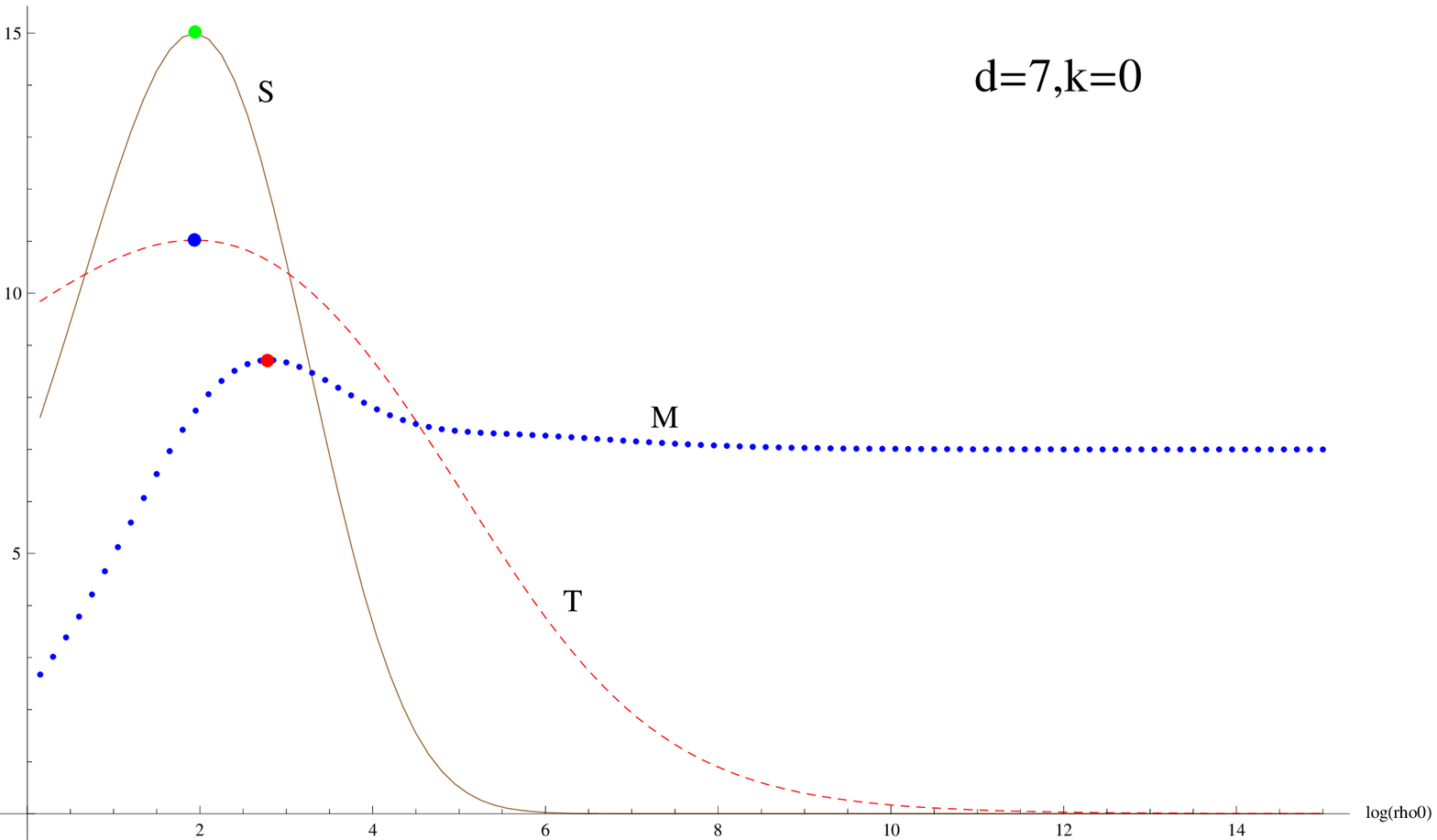}
\caption{\label{mst2d=7}The plane symmetric case: The mass
density, entropy density and the temperature of the
self-gravitating radiation configuration versus the central energy
density in the case of spacetime dimension $d=7$.}
\end{figure}

\begin{figure}
\includegraphics[width=10cm]{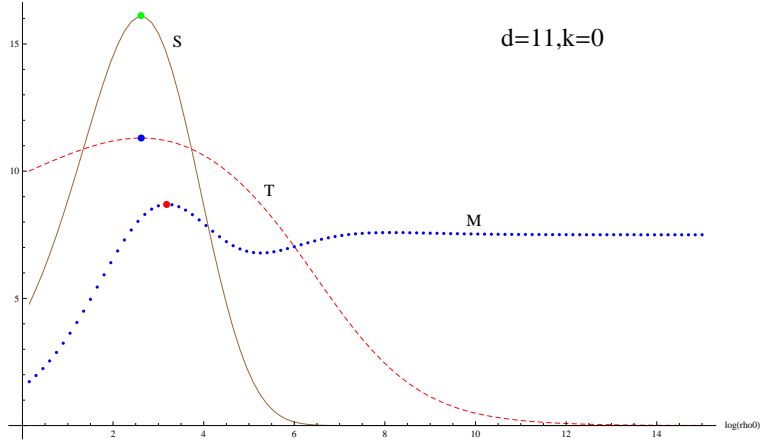}
\caption{\label{mst3d=11}The plane symmetric case: The mass
density, entropy density and the temperature of the
self-gravitating radiation configuration versus the central energy
density in the case of spacetime dimension $d=11$.}
\end{figure}

In Figs~\ref{mst1d=4}, \ref{mst2d=7} and \ref{mst3d=11} we plot
 the mass density, entropy density and temperature of
the self-gravitating radiation configuration versus the central
energy density in $d=4$, $d=7$ and $d=11$, respectively. Note that
in the figures, the mass density, entropy density and temperature
are rescaled with different scales. We observe that different from
the case of $k=1$, the maximal mass and maximal entropy
configurations do not appear at the same central energy density.
In addition, the shapes of these curves are also different, in
contrast to the case of $k=1$, where the curves of mass, entropy
and temperature are similar, see, for example, Fig.~\ref{mstk=1}
in the case of $d=4$.

Finally, let us mention that in the case of $k=-1$, there does not
exist regular equilibrium configuration of self-gravitating
radiation.  From (\ref{2eq2}) one can see that for any gravitation
mass ($m>0$), a black hole horizon always exists, there $h(r_+)=0$.
And from (\ref{2eq9}) one can see that for radiation one has $T^t_{\
t} \neq T^r_{\ r}$, therefore it is impossible to have regular
self-gravitating radiation configuration and regular black hole with
thermal equilibrium radiation with equation of state
$p=\rho/(d-1)$~\cite{LC}.

\section{Conclusion}

In this note we studied thermal equilibrium self-gravitating
radiation in higher dimensional, plane symmetric AdS space. It was
found that there exist essential differences between the spherically
symmetric configuration and plane symmetric configuration, like
their black hole solution counterparts. In particular, in the
spherically symmetric case, there is a critical dimension $d=11$: in
$ 4 \le d \le 10$, beyond the maximal mass, entropy and temperature
configurations, there is an oscillation behavior, while in $d \geq
11$, the mass, entropy and temperature become monotonic functions of
the central energy density, and approach  their asymptotic values as
the central energy density goes to infinity~\cite{Vaga,Hamm}. In the
plane symmetric case, the mass density is suppressed very much,
compared to the case of $k=1$. In each dimension, there are a
maximal mass density configuration, a maximal entropy density
configuration and a maximal temperature configuration. The maximal
mass density configuration and maximal entropy density configuration
do not appear at the same central energy density. Note that maximal
mass configuration and maximal entropy configuration do appear at
the same central energy density in the spherically symmetric case.
Although we observe there is an essential difference in the behavior
of the mass density beyond the maximal mass density configurations,
between $ 4 \le d \le 7$ and $d \geq 8$, we guess that those
configurations in the regime beyond the maximal mass density are
dynamically unstable. If so, the difference between $4 \le d \le 7$
and $d >8$ does not make any sense since it appears beyond the
maximal mass configuration and those configurations are unstable and
are of less interest physically. In addition, let us mention that
although in the case of $k=0$, there exist regular self-gravitating
radiation configurations, the Hawking-Page phase transition still
does not happen since as we have seen in the above, the contribution
of the self-gravitating radiation is negligible, compared to the
case of $k=1$. This result also can be understood from the
thermodynamic properties of the Ricci flat AdS black holes as we
discussed in the Sec.~II.

\begin{acknowledgments}
This work was supported in part by a grant from the Chinese Academy
of Sciences, grants from NSFC with No. 10325525 and No. 90403029. We
thank K. Fang for her participation in the early stage of this work.

\end{acknowledgments}

\end{document}